\documentclass{style/nseJournal}
\usepackage{amsmath}
\usepackage{amsfonts}
\usepackage{float}
\begin{document}

\title{\uppercase{Variable Dynamic Mode Decomposition for \\Estimating Time Eigenvalues in Nuclear Systems}} %title of paper

% Use the \addAuthor macro to add authors in the order they should appear. The second argument corresponds to
% the affiliation declared below.
% The corresponding author should be wrapped in \correspondingAuthor
\addAuthor{Ethan Smith}{a}
% The corresponding author's email can be specified using \correspondingEmail
\addAuthor{Ilham Variansyah}{a}
\addAuthor{\correspondingAuthor{Ryan McClarren}}{a}
\correspondingEmail{rmcclarr@nd.edu}
% Affiliations can be added in the order they should appear. For breaks in addresses, use either \\ or \tabularnewline
\addAffiliation{a}{Department of Aerospace and Mechanical Engineering\\ University of Notre Dame\\ Fitzpatrick Hall, Notre Dame, IN 46556}

% Add keywords to appear in Abstract in the order they should appear
\addKeyword{time eigenvalues}
\addKeyword{data-driven algorithms}
\addKeyword{time-dependent neutronics}

\titlePage

\begin{abstract}
We present a new approach to calculating time eigenvalues of the neutron transport operator (also known as $\alpha$ eigenvalues) by extending the dynamic mode decomposition (DMD) to allow for non-uniform time steps. The new method, called variable dynamic mode decomposition (VDMD), is shown to be accurate when computing eigenvalues for systems that were infeasible with DMD due to a large separation in time scales (such as those that occur in delayed supercritical systems). The $\alpha$ eigenvalues of an infinite medium neutron transport problem with delayed neutrons and consequently having multiple, very different relevant time scales are computed. Furthermore, VDMD is shown to be of similar accuracy to the original DMD approach when computing eigenvalues in other systems where the previously studied DMD approach can be used.
\end{abstract}

\section{Introduction}
The time behavior of neutron transport problems can be understood through the spectrum of the transport operator given by 
time eigenvalues \cite{Bell_Glasstone}, also known as $\alpha$ eigenvalues. The corresponding eigenfunctions for the time eigenvalues are angular fluxes and, therefore, are more challenging to compute than the more well-known $k$ eigenvalue problem where the eigenfunctions are scalar fluxes. 

The conventional approach to computing $\alpha$ eigenvalues is via the $k$-$\alpha$ iteration procedure \cite{Brockway,Variansyah}. This method is ill-suited for many subcritical problems because the iteration procedure can introduce total cross-sections that are effectively negative. Moreover, $k$-$\alpha$ iterations can only calculate the eigenvalue farthest to the right in the complex plane. In a given scenario, especially at short times, other eigenvalues may be more important to the dynamics.

The dynamic mode decomposition (DMD) has been used in the neutron transport community for the purpose of estimating time eigenvalues in calculations \cite{McClarren} and experiments \cite{yamamoto2021application,yamamoto2022higher}, as well as to create reduced-order models \cite{hardy2019dynamic}, improve the convergence of power iteration in $k$-eigenvalue problems \cite{roberts2019acceleration,yamamoto2022dynamic}, and to accelerate source iteration \cite{MCCLARREN2022110756,mcclarren2018acceleration}. Originally introduced for fluid dynamics problems \cite{schmid2010dynamic,tu2013dynamic}, DMD takes the solution to a time-dependent transport problem and uses the solution at several time steps to estimate an approximate transport operator based solely on the available solutions. This approximate operator has eigenvalue-eigenvector pairs that are also eigenvalues and eigenvectors of the full transport operator. DMD is a fully data-driven method; this means that it will find the eigenmodes that are important in the system's evolution. Additionally, the approximate operator that DMD estimates can be used as a reduced order model to evolve the system forward in time.

%DMD has also found applications in other transport problems. These include accelerating the convergence of source iteration in discrete ordinates calculations \cite{McClarren_Haut}, accelerating power iterations in $k$-eigenvalue problems \cite{Roberts}, and accelerating problems where there are nonlinear positivity fixes in the iteration strategy \cite{McClarren_Haut_nonlinear}.

The DMD procedure requires that data snapshots be spaced at regular intervals (i.e., have the same time step between the snapshots). This is a significant drawback for transport problems with delayed neutrons or problems very near critical, as we will demonstrate. One would prefer to use variable time steps to resolve important transients without losing efficiency. This was the motivation for developing a variable DMD algorithm that allows for irregularly spaced steps. In this paper we develop this method and show how it can be used for a variety of time integration techniques. The underlying idea is the same as in \cite{McClarren}: use solutions to a time-dependent transport problem to estimate the $\alpha$ eigenvalues of a system. We demonstrate on a variety of problems that our method can accurately estimate the eigenvalues of the system without the requirement of equal-sized time steps. One restriction of our approach is that it does require knowledge of the type of time discretization used in producing the solution.

%One drawback of DMD is that it requires the solution data to be at equally spaced time intervals. This drawback is significant for transport problems with delayed neutrons as we will demonstrate. It would be better to allow the time step to increase after the initial prompt transients decay away. Such a variable time step, perhaps logarithmically spaced, would be more efficient.

%In this paper, we present a dynamic mode decomposition method that allows for variable time steps that we call variable dynamic mode decomposition VDMD. This method changes the DMD procedure to allow for variable time steps, but does need to know the type of time discretization used in producing the solution.  In the following section we present the method and then show how it can be used on a variety of problems.

\section{\uppercase{Theory of a Variable Step Decomposition} }
\label{sec:first}
We extend the DMD procedure to accept a variable time step for determining $\alpha$ eigenvalues by leveraging the fact that we know which time discretization method was used to create the data matrix. We begin with a generic, linear system of differential equations of the form\footnote{The DMD method, and our extension, can be applied to nonlinear problems. However, because we are interested in neutron transport problems primarily, beginning from a linear system is natural.}
\begin{equation}
    \label{ODE}
    \frac{d}{d t} \mathbf{y}(t) = \mathbf{Ay}(t),
\end{equation}
with initial conditions $\mathbf{y}(0) = \mathbf{y}^0$. The vector $\mathbf{y}$ is of length $M$, and the matrix $\mathbf{A}$ is of size $M \times M$. In neutron transport applications $\mathbf{A}$ represents a discretized transport operator, but for most neutron transport calculations this matrix is never explicitly formed. The vector $\mathbf{y}$ contains the spatial, angular, and energy degrees of freedom in the solution.

%For the problems presented in this paper we have used the backward Euler method, which conveniently relates our matrix operator $A$ and sequential data vectors $\mathbf{y}^n$ and $\mathbf{y}^{n+1}$ via
%\begin{equation}
%  \label{BEVDMD}
%  \frac{\mathbf{y}^{n+1} - \mathbf{y}^{n}}{t^{n+1}- t^n} = \mathbf{A y}^{n+1},
%\end{equation}

%where the superscripts denote a time level and $t^n$ is the time at the beginning of the $n$th step. 
To develop our variable DMD method, we consider common, implicit time integration procedures for Eq.~\eqref{ODE}. The backward Euler method applied to this equation is
\begin{flalign}
    &  & \frac{\mathbf{y}^{n+1} - \mathbf{y}^n}{t^{n+1}-t^{n}} &= \mathbf{A} \mathbf{y}^{n+1}. & & \text{\llap{Backward Euler}}\label{eq:BE} 
    \end{flalign}
Here the superscripts denote a time step number: $\mathbf{y}^n \approx \mathbf{y}(t^n)$ is the approximation of the solution at time $t^n$. The time step size is the difference between $t^{n+1}$ and $t^n$. Commonly, this equation would be factored to show how to produce $\mathbf{y}^{n+1}$ from $\mathbf{y}^n$. However, the form in Eq.~\eqref{eq:BE} will be most useful to us in developing a DMD-like procedure for approximating the operator $\mathbf{A}$. 

The Crank-Nicolson \cite{mcclarren2017computational} and BDF-2 \cite[Chap.~V.1]{wanner1996solving} methods applied to Eq.~\eqref{ODE} yield the following equations:
\begin{flalign}
    &  & \frac{\mathbf{y}^{n+1} - \mathbf{y}^n}{t^{n+1}-t^{n}} &= \frac{1}{2}\mathbf{A} (\mathbf{y}^{n+1}+\mathbf{y}^{n}). & & \text{\llap{Crank-Nicolson}}\label{eq:CN} \\
    &  & \frac{\mathbf{y}^{n+1} - \frac{4}{3}\mathbf{y}^{n} + \frac{1}{3}\mathbf{y}^{n-1}}{\frac{2}{3} (t^{n+1} - t^{n} )} &= \mathbf{A} \mathbf{y}^{n+1}. & & \text{\llap{BDF-2}}\label{eq:BDF}
    \end{flalign}
For each of these methods we can write the update as 
\begin{equation}
    \mathbf{u}^{n+1} = \mathbf{A}\mathbf{v}^{n},
\end{equation}
where 
\begin{equation}
    \mathbf{u}^{n+1} = \frac{1}{t^{n+1} - t^n} \begin{cases} \mathbf{y}^{n+1} - \mathbf{y}^n & \text{Backward Euler or Crank-Nicolson}\\
    \frac{3}{2}\left(\mathbf{y}^{n+1} - \frac{4}{3}\mathbf{y}^{n} + \frac{1}{3}\mathbf{y}^{n-1}\right) & \text{BDF-2}
    \end{cases},
\end{equation}
and
\begin{equation}
    \mathbf{v}^{n} = \begin{cases} \mathbf{y}^{n+1} & \text{Backward Euler or BDF-2}\\
    \frac{1}{2}\left(\mathbf{y}^{n+1} + \mathbf{y}^{n} \right) & \text{Crank-Nicolson}
    \end{cases}.
\end{equation}
If we consider the repeated application of the time integration method from time $t^0 = 0$ over $N$ time steps, we could collect via concatenation the vectors $\mathbf{u}^{n}$ and $\mathbf{u}^{n+1}$ into matrices of size $M \times N$, $\mathbf{U}^{+}$ and $\mathbf{V}^-$ defined as
\begin{equation}\label{eq:uandv}
    \mathbf{U}^+ = \left[\mathbf{u}^N\, \mathbf{u}^{N-1}\, \dots\, \mathbf{u}^1\right], \qquad  \mathbf{V}^- = \left[\mathbf{v}^{N-1}\, \mathbf{v}^{N-2}\, \dots\, \mathbf{v}^0\right].
\end{equation}
Using these definitions we can write
\begin{equation}\label{eq:overall_relation}
    \mathbf{U}^+ = \mathbf{A} \mathbf{V}^-.
\end{equation}
We note that BDF-2 is not self-starting because it needs the two previous solutions. Either backward Euler or Crank-Nicolson can be used to compute $\mathbf{u}^1$ from the initial condition and still fit the form we have here. In our formulation, the operator $\mathbf{A}$ does not depend on time so that the time step size does not affect it; all of the information regarding the time step size is contained in vectors $\mathbf{u}$ and $\mathbf{v}$. 

We now proceed as is standard for DMD. We first take the thin singular value decomposition (SVD) of $\mathbf{V}^{-}$ and write this as
\begin{equation}
    \mathbf{V}^{-} = \mathbf{L} \mathbf{S} \mathbf{R}^\mathrm{T},
\end{equation}
where $\mathbf{L}$ is of size $M\times r$, $\mathbf{S}$ is a diagonal matrix of size $r \times r$ with positive entries called singular values, and $\mathbf{R}$ is of size $N \times r$ where $r$ is number of non-zero singular values (or singular values with a magnitude larger than some threshold). The matrices $\mathbf{L}$ and $\mathbf{R}$ have the property that
\begin{equation}
    \mathbf{L}^\mathrm{T} \mathbf{L} = \mathbf{I}_M, \qquad \mathbf{R}^\mathrm{T} \mathbf{R} = \mathbf{I}_N,
\end{equation}
where $\mathbf{I}_K$ is an identity matrix of size $K \times K$. Using this property, we right multiply Eq.~\eqref{eq:overall_relation} by $\mathbf{R}\mathbf{S}^{-1}$ and then left multiply by $\mathbf{L}^\mathrm{T}$ to get 
\begin{equation}\label{eq:approxA}
\mathbf{L}^\mathrm{T} \mathbf{U}^+ \mathbf{R}\mathbf{S}^{-1} = \mathbf{L}^\mathrm{T} \mathbf{A L} \equiv \tilde{\mathbf{A}}. 
\end{equation}
The matrix $\tilde{\mathbf{A}}$ is an $r \times r$ approximation to the operator $\mathbf{A}$. Moreover, we can compute this approximation using only the known data matrices $\mathbf{U}^+$ and $\mathbf{V}^-$ as indicated by the LHS of Eq.~\eqref{eq:approxA}. 

As shown in previous work \cite{McClarren,tu2013dynamic}, the eigenvalues of $\mathbf{L}^{T} \mathbf{A L}$ are also eigenvalues of $A$, and if $\mathbf{w}$ is an eigenvector of $\mathbf{L}^{T} \mathbf{A L}$, then $\mathbf{Lw}$ is an eigenvector of $\mathbf{A}$. These properties allow us to estimate eigenvalues/eigenvectors of $\mathbf{A}$ without any knowledge of the operator itself, other than the results of calculating solutions using one of the time discretization schemes above.

Equation \eqref{eq:approxA} has the same form as the standard DMD approximation, except we have arrived at it using different data matrices. We call this approach the variable dynamic mode decomposition (VDMD). One key difference between VDMD and the standard DMD method is that we require knowledge of how the solution is updated because the time integration method and the time step sizes influence the matrices $\mathbf{U}^+$ and $\mathbf{V}^-$. This means that we cannot apply this method directly to experimental measurements without having an approximation of the time derivative of the measurement. Further investigation of the application of this approach to measured data should be the subject of future research.

\subsection{Demonstration on a simple problem}
Consider the system
\begin{equation}\label{eq:damped_system}
    \frac{\partial}{\partial t}\begin{bmatrix}
        y_1 \\ y_2
    \end{bmatrix} = {\begin{bmatrix}
       0 & 1 \\
        -\omega^2 & -\lambda 
    \end{bmatrix}}
    \begin{bmatrix}
        y_1 \\ y_2
    \end{bmatrix} = \mathbf{A} \begin{bmatrix}
        y_1 \\ y_2
    \end{bmatrix},
\end{equation}
with initial conditions $y_1(0) = 1$ and $y_2(0) = 0$. This system is the first-order system corresponding to the second-order ODE
\begin{equation}
    y_1'' + \lambda y_1' + \omega^2 y_1 = 0, \qquad y_1(0) = 1,\quad y_1'(0) = 0
\end{equation}
with solution
\begin{equation}
    y_1(t) = \left(\cos( w t) + \frac{\sin(w t)}{w \tau}\right)e^{-t/\tau},
\end{equation}
where
\begin{equation}
    \tau = \frac{2}{\lambda}, \qquad w^2 = \omega^2 - \frac{\lambda^2}{4}.
\end{equation}
For this problem, the matrix $\mathbf{A}$ has eigenvalues given by $ \left(-\lambda\pm \sqrt{\lambda ^2-4 \omega ^2} \right)/2$. 

We will solve system \eqref{eq:damped_system} using the various time integrators mentioned above. Setting $\lambda = 1/10$ and $\omega = 13 \sqrt{29}/20 \approx 3.500357$, we can analytically determine the eigenvalues of $\mathbf{A}$ to be $0.05 \pm 3.5 i$. To demonstrate the VDMD method we solve this problem with 20 logarithmically spaced time steps starting from $10^{-3}$ and increasing to a final step size of 3.0 (i.e., the step size increases by a factor of about 1.524 each step). 

\begin{figure}
\includegraphics[]{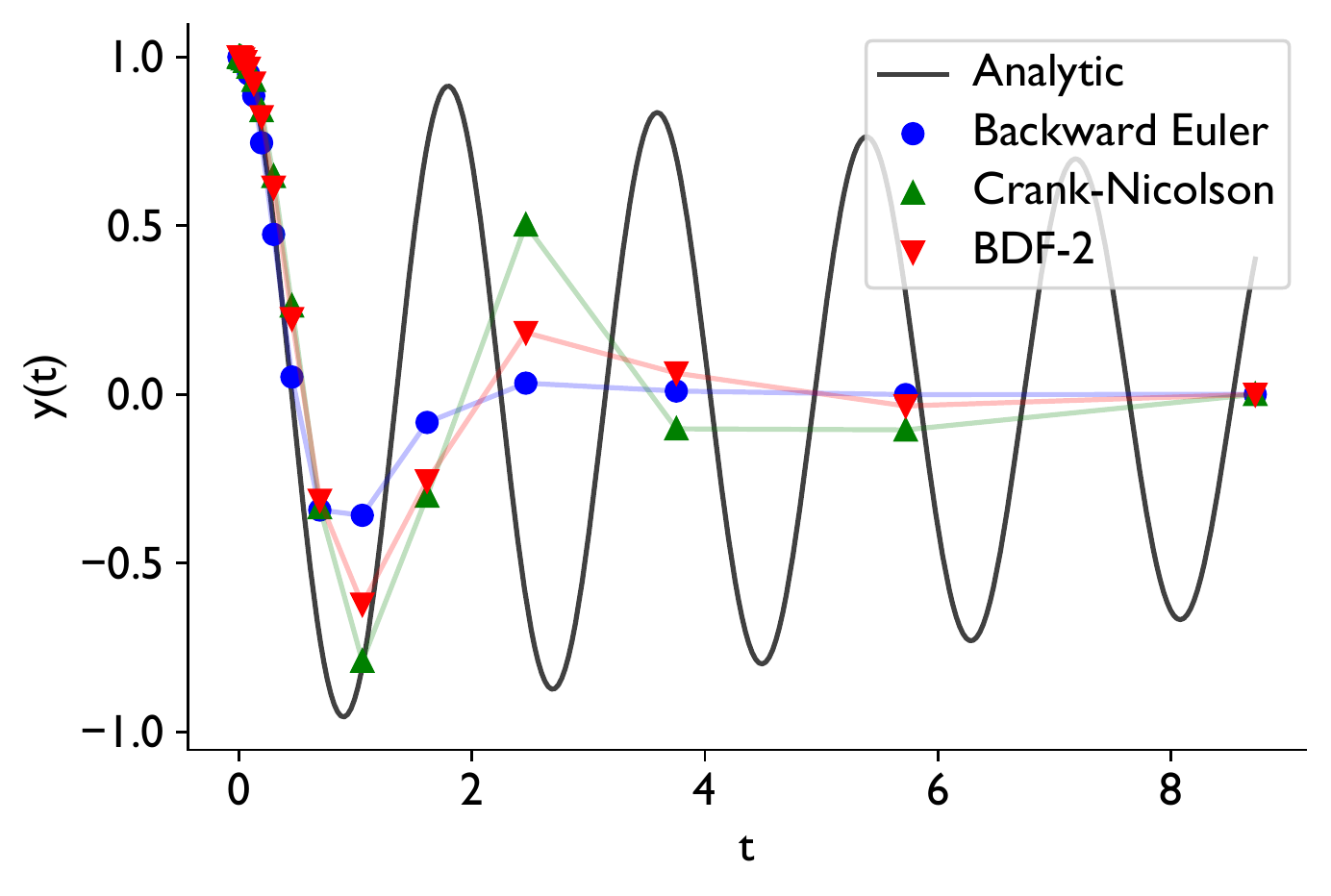}
\caption{Data used to estimate the eigenvalues of the damped oscillator problem using VDMD. The solutions shown here used 20 logarithmically spaced time steps where the first step was of size 0.001 and the final step was of size 3. Despite the numerical error, VDMD is able to estimate the eigenvalues of the underlying operator to machine precision.}\label{fig:dampedwave}
\end{figure}

As can be seen from Figure \ref{fig:dampedwave} the various numerical solutions to this problem do not agree with the analytic solution with these large time step sizes. Nevertheless, applying the VDMD method to these numerical solutions produces eigenvalues that agree with the exact values to within machine precision (i.e., relative errors on the order of $10^{-14}$).  Other numerical experiments indicate that with as few as 3 time steps we can approximate the eigenvalues to machine precision. This is an important result because it indicates that VDMD, because it has knowledge of the numerical method used to approximate the solution, is able to estimate properties of $\mathbf{A}$ even when the numerical solutions in the snapshot matrices have large amounts of time discretization error.

\section{VDMD and Time Eigenvalues for Neutronics} 
\label{sec:second}
Now that we have presented the VDMD method in general, we turn to the neutron transport problem. We begin with the time-dependent transport equation including delayed neutrons \cite{Bell_Glasstone}
\begin{equation} \label{eq:tdtransport}
\frac{\partial \psi}{\partial t}  = \mathcal{A} \psi + \sum_{i=1}^I \frac{\chi_{di}(E)}{4\pi} \lambda_i C_i(x,t) ,
\end{equation}
\begin{equation} \label{eq:tdtransport_delayed}
\frac{\partial C_i}{\partial t}  = \int_{0}^\infty dE \, \beta_i\nu\Sigma_f(E) \phi(x,E,t) - \lambda_i C_i(x,t), \qquad i=1,\dots,I,
\end{equation}
where $\psi(x,\Omega,E,t)$ is the angular flux at position $x \in \mathbb{R}^3$, in direction $\Omega \in \mathbb{S}_2$, at energy $E$ and time $t$ and $C_i(x,t)$ is the delayed precursor density of flavor $i$. The transport operator $\mathcal{A}$ is given by
\[ \mathcal{A} = v(E)(-\Omega \cdot \nabla -\Sigma_t + \mathcal{S} + \mathcal{F}),\]
with $\mathcal{S}$ and $\mathcal{F}$ the scattering and fission operators:
\begin{equation}\label{eq:scatt}
\mathcal{S} \psi = \int_{4\pi} d\Omega'  \int_{0}^\infty dE' \, \Sigma_s(\Omega' \rightarrow \Omega, E' \rightarrow E) \psi(x,\Omega',E',t), 
\end{equation}
\begin{equation}\label{eq:fiss}
\mathcal{F} \psi = \frac{\chi_p(E)}{4\pi}\int_{0}^\infty dE' \, (1-\beta)\nu\Sigma_f(E') \phi(x,E',t), 
\end{equation}
where $\Sigma_s(\Omega' \rightarrow \Omega, E' \rightarrow E) $ is the double-differential scattering cross-section from direction $\Omega'$ and energy $E'$ to direction $\Omega$ and energy $E$ , $\nu\Sigma_f(E')$ is the fission cross-section times the expected number of fission neutrons at energy $E'$, and $\chi_p(E)$ is the probability of a fission neutron being emitted with energy $E$, $\chi_{di}(E)$ is the probability of a delayed neutron of flavor $i$ being emitted with energy $E$, $\beta_i$ is the fraction of fission neutrons that come from delayed flavor $i$, $\sum_i \beta_i = \beta$, and $\lambda_i$ is the decay constant for precursor flavor $i$. The scalar flux $\phi(x,E,t)$ is defined as the integral of the angular flux over the unit sphere,
\begin{equation}\label{eq:phi}
\phi(x,E,t) = \int_{4\pi} d\Omega\, \psi(x,\Omega,E,t).
\end{equation}

We study the use of VDMD on the transport equation with discretizations of the multigroup method \cite{Bell_Glasstone} in energy, discrete ordinates in angle, and various spatial discretizations.  To connect with Eq.~\eqref{ODE}, the time-dependent transport equation can be written as a system of differential equations
\begin{equation} \label{eq:tdtransport_dsic}
\frac{\partial \Psi}{\partial t}  = \mathbf{A} \Psi(t),
\end{equation}
where $\Psi(t)$ is a time-dependent vector of the discrete values of the angular flux $\psi$ at each space, energy, and angle degree of freedom and the delayed neutron precursor densities at each spatial degree of freedom. The discrete transport/delayed neutron operator matrix is written as $\mathbf{A}$.  Notice that the time-eigenvalue problem which supposes a form for $\Psi(t)$ of $e^{\alpha t} \Psi_\alpha$ leads to the eigenvalue problem
\begin{equation} \label{eq:tdtransport_dsic_eval}
\alpha \Psi_\alpha  = \mathbf{A} \Psi_\alpha.
\end{equation}
From this form we can directly apply the VDMD method as detailed above:
\begin{enumerate}
    \item Take $N$ backward Euler (or Crank-Nicolson or BDF-2) steps of the discrete transport problem and form the matrices $\mathbf{U}^+$ and $\mathbf{V}^-$ [Eq.~\eqref{eq:uandv}].
    \item Compute the SVD of $\mathbf{V}^-$ and form $\Tilde{\mathbf{A}}$ [Eq. \eqref{eq:approxA}].
    \item Compute the eigenvalues of $\Tilde{\mathbf{A}}$ and the associated eigenvectors.
\end{enumerate}

Previously published versions of DMD for estimating time eigenvalues \cite{McClarren} required a fixed-sized time step to generate the data matrices. VDMD does not have this constraint, though it does require the knowledge of the time integration method used.   The ability to handle variable-sized time steps is an important advance because, for problems where delayed neutrons are significant, computing  the prompt and delayed modes of the system would require minuscule steps to capture the prompt scales and a large number of these steps to integrate to times when delayed neutrons are significant, as we demonstrate below.

The algorithm has flexibility in that the initial condition for the time-dependent calculation can be chosen based on the analysis being performed. For example, to model an experiment and extract the important eigenmodes, one would initialize the problem with the appropriate experimental initial condition. Alternatively, to compute the dominant eigenmodes in a system, one could use a random initial condition to assure that all of the modes are excited in the system.
\section{NUMERICAL RESULTS}
\subsection{Infinite Medium Problem with Delayed Neutrons}
%We begin with a the neutron flux solution of a twelve group problem with six delayed neutron precursor groups and a buckling approximation to model a finite-sized sphere. This problem will demonstrate the need for variable time steps in the VDMD process. For this problem the transport operator is an $18 \times 18$ matrix corresponding to the 12 groups and 6 precursor types. We can directly estimate the eigenvalues of this matrix using standard linear algebra techniques. The eigenvalues change as the radius in the buckling is adjusted; this allows us to study sub- and delayed supercritical configurations.  

To demonstrate the need for a variable time step method, we consider the neutron flux solution of a twelve group problem with six delayed neutron precursor groups and a spherical buckling approximation for the leakage. This problem results in a transport operator that is an $18 \times 18$ matrix for which we can use numerical linear algebra software to estimate the eigenvalues. A subcritical and  delayed supercritical case are considered by modifying the radius of the sphere in the buckling. %Since the problem is essentially an infinite medium problem, the matrix transport operator $A$ can be exactly obtainted, and reference $\alpha$ eigenvalues can be computed using linear algebra to compare against the $\alpha$ eigenvalue results of the VDMD technique. 
The numerical solution is computed using both  backward Euler and Crank-Nicolson time integration, and  logarithmically-spaced timesteps are used.

The systems considered had a radius of $11.7335$ cm for the subcritical case and a radius of $11.735$ cm for the supercritical case. These unsteady problems were initialized with a single neutron per cm$^{3}$ in the highest energy group at time zero (to approximate probing the system with a fast-neutron source), and used a logarithmically spaced time grid to a final time of $10^{3}$ s, using $200$ time steps; the first time step is of size $\Delta t = 10^{-11}$ s. A challenge of using the traditional DMD algorithm to analyze these kinds of problems lies in the very different time scales over which the features of the system manifest. As an example, the prompt multiplication peak occurs around ten nanoseconds into the problem, while the delayed multiplication continues until well into the hundreds of seconds. This is most evident in Fig.~\ref{fig:inf_log_sup}, where the supercritical behavior of the system is not evident until over one hundred seconds into the evolution of the problem.

\begin{figure}[H]
  \centering
    \includegraphics[scale=0.8]{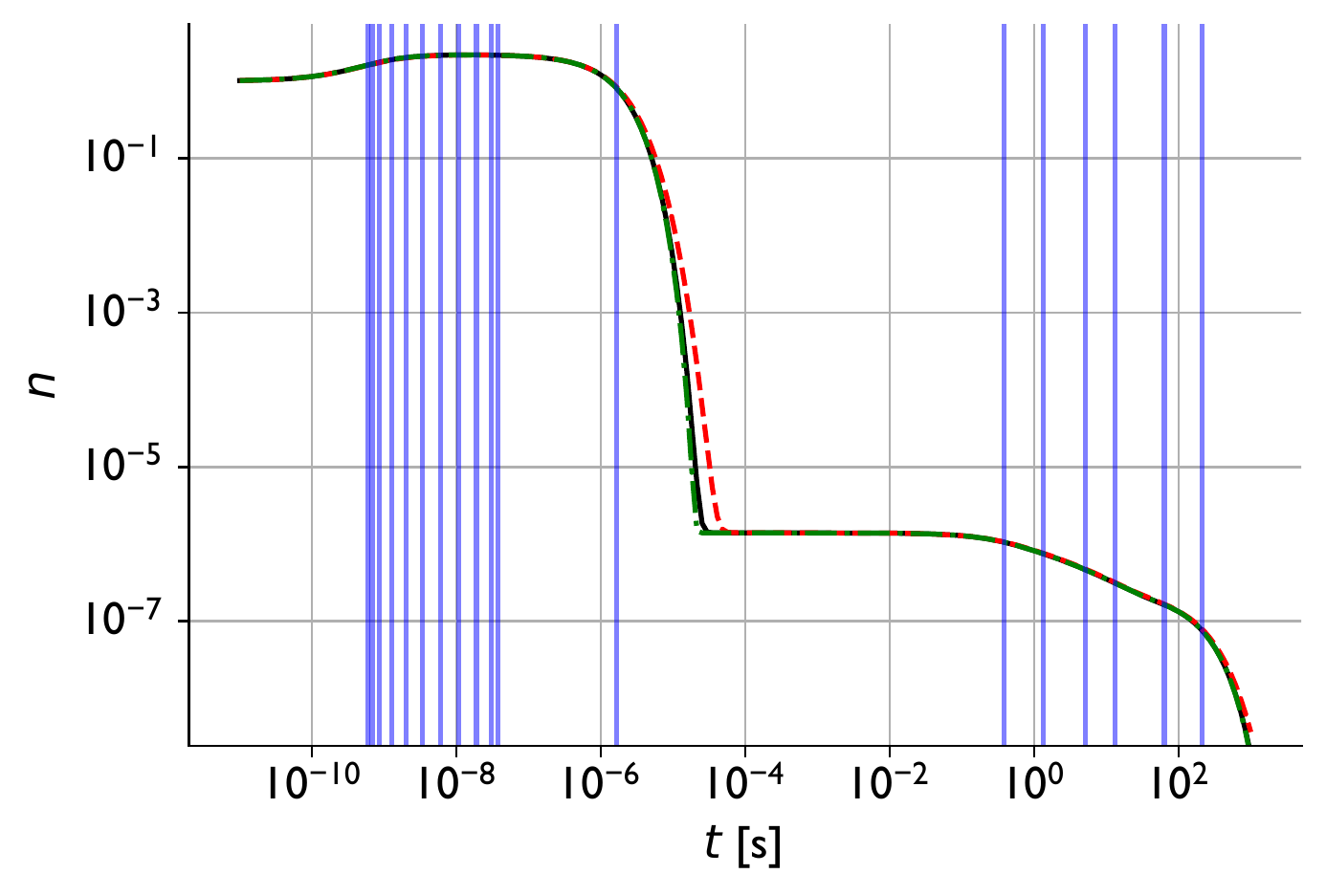}
   \caption{Solution in terms of the number density of neutrons, $n(t)=\phi(t)/v(t)$, for the subcritical sphere problem. The black line plots the analytical solution, the red dashed line plots the backward Euler solution, and the green dash-dot line is the Crank-Nicolson solution. Vertical blue lines are negative eigenperiods, the inverse of negative part of the $\alpha$ eigenvalues. }
   \label{fig:inf_log_sub}
\end{figure}

\begin{figure}[H]
  \centering
    \includegraphics[scale=0.8]{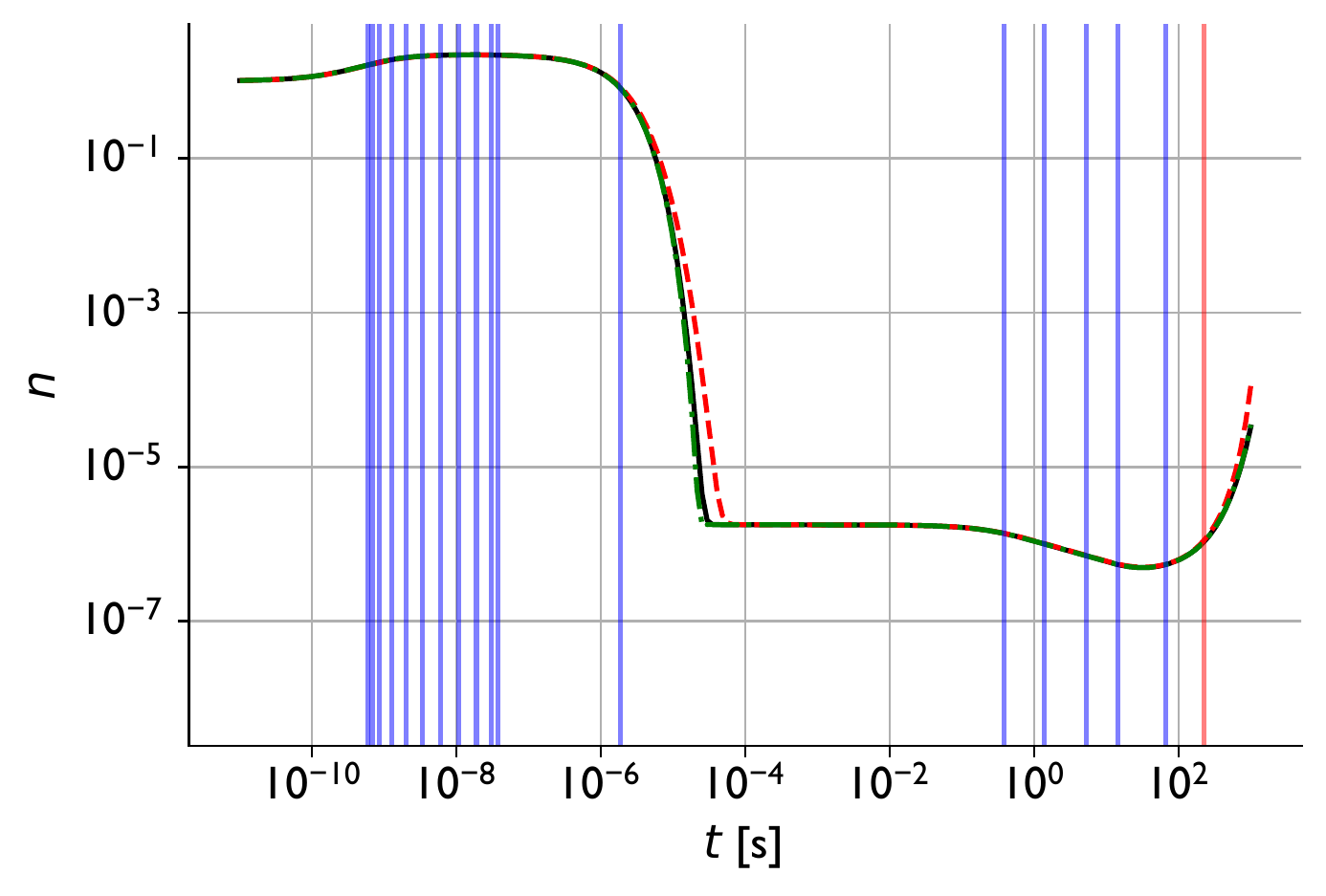}
   \caption{Solution in terms of the number density of neutrons, $n(t)$, for the delayed supercritical sphere problem. The labeling is the identical to Figure \ref{fig:inf_log_sub} with the addition of a vertical red line to denote the positive eigenperiod.}
   \label{fig:inf_log_sup}
\end{figure}

Figures \ref{fig:inf_log_sub} and \ref{fig:inf_log_sup} plot the neutron population in the sphere over time. Resolving the early and late time features of this delayed subcritical system would be extremely expensive  using a uniform timestep, the present example uses just $200$ time points. In these figures the black line corresponds to the exact solution based on the matrix exponential, the red dashed line is the backward Euler solution, and the green dash-dot line is the Crank-Nicolson solution. The vertical blue lines correspond to the magnitude of the periods of the exact eigenvalues ( $\left|\alpha\right|^{-1}$), where blue and red colors respectively indicate negative and positive value. 
We also note that the Crank-Nicoloson solution is closer to the analytic solution obtained using a matrix exponential. The biggest errors in the numerical solutions occur during the rapid transition between $10^{-6}$ and $10^{-4}$ seconds. 

The exact eigenvalues, along with the relative error in pcm (1 pcm $= 10^{-5}$) for the eigenvalues estimated with VDMD and either backward Euler or Crank-Nicolson, are tabulated for both the subcritical and supercritical cases in Table~\ref{tab:VDMD_EIG}. In the table we can see that for all 18 eigenvalues the VDMD estimates are within 1 pcm of the reference. We also observe that there does not appear to be a clear benefit to using Crank-Nicolson over backward Euler, despite the fact that the Crank-Nicolson is a second-order accurate method. This speaks to part of the benefit of VDMD: VDMD is aware of the discretization so it can reasonably approximate the operator that generated the data.

\begin{table}[H]
    \centering
    \caption{\bf VDMD Eigenvalues Errors in pcm for Subcritical and Delayed Supercritical Spheres using Backward Euler (BE) and Crank-Nicolson (CN)}
    \label{tab:VDMD_EIG}
\begin{tabular}{lll||lll}
\multicolumn{3}{c||}{Subcritical}                                                         & \multicolumn{3}{c}{Supercritical}  \\
\hline
Analytic (s$^{-1}$) & BE Error & CN Error & Analytic (s$^-1$) & BE Error & CN Error\\
\hline
      -1.67621$\times 10^{9}$ & 4.12074$\times 10^{-7}$ & 9.15307$\times 10^{-7}$ &      -1.67597$\times 10^{9}$ & 3.5385$\times 10^{-7}$  & 3.48643$\times 10^{-7}$ \\
      -1.46266$\times 10^{9}$ & 1.04743$\times 10^{-6}$ & 2.41534$\times 10^{-6}$ &      -1.46245$\times 10^{9}$ & 8.39863$\times 10^{-7}$ & 8.98227$\times 10^{-7}$ \\
      -1.17322$\times 10^{9}$ & 7.94049$\times 10^{-7}$ & 2.67553$\times 10^{-6}$ &      -1.17306$\times 10^{9}$ & 1.38875$\times 10^{-6}$ & 9.82629$\times 10^{-7}$ \\
      -7.9423$\times 10^{8}$  & 1.40458$\times 10^{-7}$ & 1.9733$\times 10^{-6}$  &      -7.94119$\times 10^{8}$ & 3.57139$\times 10^{-7}$ & 7.95791$\times 10^{-7}$ \\
      -4.99878$\times 10^{8}$ & 5.88453$\times 10^{-7}$ & 1.61807$\times 10^{-6}$ &      -4.99804$\times 10^{8}$ & 8.25479$\times 10^{-7}$ & 5.93383$\times 10^{-7}$ \\
      -2.92659$\times 10^{8}$ & 4.74705$\times 10^{-7}$ & 1.42446$\times 10^{-6}$ &      -2.92606$\times 10^{8}$ & 1.03237$\times 10^{-7}$ & 5.74564$\times 10^{-7}$ \\
      -1.66397$\times 10^{8}$ & 1.48477$\times 10^{-8}$ & 1.43941$\times 10^{-6}$ &      -1.66365$\times 10^{8}$ & 2.66235$\times 10^{-7}$ & 5.597$\times 10^{-7}$   \\
      -9.36139$\times 10^{7}$ & 9.43281$\times 10^{-8}$ & 1.49989$\times 10^{-6}$ &      -9.35967$\times 10^{7}$ & 5.99602$\times 10^{-7}$ & 5.64083$\times 10^{-7}$ \\
      -5.28136$\times 10^{7}$ & 2.51533$\times 10^{-7}$ & 1.89254$\times 10^{-6}$ &      -5.28057$\times 10^{7}$ & 3.02384$\times 10^{-6}$ & 5.04243$\times 10^{-7}$ \\
      -3.3247$\times 10^{7}$  & 1.57121$\times 10^{-6}$ & 3.51607$\times 10^{-6}$ &      -3.32436$\times 10^{7}$ & 5.85692$\times 10^{-6}$ & 1.14324$\times 10^{-6}$ \\
      -2.64094$\times 10^{7}$ & 3.76945$\times 10^{-5}$ & 1.03566$\times 10^{-5}$ &      -2.64085$\times 10^{7}$ & 0.000125423 & 9.61842$\times 10^{-6}$ \\
 -609650           & 2.07949$\times 10^{-8}$ & 5.25697$\times 10^{-8}$ & -540139           & 9.785$\times 10^{-9}$   & 9.24618$\times 10^{-8}$ \\
      -2.6143      & 0.0002196   & 0.0015871   &      -2.60143     & 0.000776098 & 0.00106403  \\
      -0.743522    & 0.00158942  & 0.00683878  &      -0.732142    & 0.00555557  & 0.00462957  \\
      -0.196729    & 0.0077539   & 0.0220177   &      -0.187957    & 0.0212489   & 0.0141573   \\
      -0.0766105   & 0.0123597   & 0.0339459   &      -0.0699507   & 0.0301533   & 0.0208952   \\
      -0.0158092   & 0.0221887   & 0.0445097   &      -0.0149391   & 0.0202711   & 0.018085    \\
      -0.00467553  & 0.0408661   & 0.0300762   &       0.00441678  & 0.00430661  & 0.000104375 \\
\hline
\end{tabular}
\end{table}

\subsection{Modak and Gupta Problem}
To demonstrate that VDMD is also applicable to numerical solutions to transport problems, and to compare it with the original DMD formulation, we consider a transport problem first published by Modak and Gupta \cite{Modak&Gupta} with semi-analytic results published by Kornreich and Parsons \cite{kornreich2005time}. Unlike the infinite medium problem, the exact time evolution operator cannot be explicitly formed. The problem is defined as a 10 mean-free-path, non-multiplying slab of two materials, with $\Sigma_S = 10$ in one material and $\Sigma_S = 9$ in the other. The other nuclear properties are defined as $\Sigma_T = 10$, $\nu\Sigma_f = 0$, and $\chi = 0$. The grain size is varied for separate runs of this simulation, this parameter defines the fraction of the overall length that each slice of material is, compared to the whole. These material slices are arranged in alternating order. The zero grain size case indicates a homogeneous material with average properties. Given that we did not see a large benefit in using Crank-Nicolson in the previous problem, in this problem we exclusively use backward Euler time discretization. 

Previous work has demonstrated that to accurately estimate eigenvalues on these problems, high resolution in space and angle is necessary. The Modak and Gupta problem was solved with VDMD using a $S_{196}$ discrete ordinates approach with $1000$ spatial grid points and 101 logarithmically spaced time steps with the first time step ending at $t = 10^{-5}$ and the last ending at $t = 100$; results using the original, equi-spaced time step formulation of DMD used 101 equally spaced time steps over a time range of $100$ using the same spatial and angular discretizations. 

We compare the results of the new VDMD algorithm and the existing DMD algorithm. The problem is initialized with a random initial condition for the angular flux in all angles and all positions sampled uniformly between zero and one, to ensure that all eigenmodes are excited.  The four largest real eigenvalues from these numerical experiments are tabulated in Table \ref{tab:MandG}. Bold numbers indicate an agreement for the number in that decimal place with the published  results \cite{kornreich2005time}. %We note that in some cases DMD and VDMD identify other, more negative real eigenvalues, but these are not shown in the table because we do not have a reference calculation for them.
From the table we notice that VDMD and DMD have nearly identical performance in the number of digits matched with the semi-analytic results. This demonstrates that we have not sacrificed accuracy in formulating a variable step version of DMD.

\begin{table}[H]
    \centering
    \caption{\bf Computed eigenvalues for the Modak and Gupta problem using VDMD and DMD compared with the semi-analytic results.}\label{tab:MandG}
\begin{tabular}{c||ccc}
\hline
Grain Size & Semi-Analytic & VDMD      & DMD       \\ \hline
        & -0.551429       & \textbf{-0.55}0814 & \textbf{-0.55}0814 \\ %\hline
0.5           & -1.71149        & \textbf{-1.7}0646  & \textbf{-1.7}0645  \\ %\hline
           & -2.94399        & \textbf{-2.94}235  & \textbf{-2.94}235  \\ %\hline
           & -5.28234        & \textbf{-5}.16961  & \textbf{-5}.17338  \\ \hline
%           &                 &           &           \\
\hline
       & -0.703578       & \textbf{-0.70}4010 & \textbf{-0.70}4010 \\ %\hline
 0.25          & -1.45315        & \textbf{-1.45}044  & \textbf{-1.45}044  \\ %\hline
           & -3.07282        & \textbf{-3.07}073  & \textbf{-3.07}073  \\ %\hline
           & -5.26925        & \textbf{-5}.15114  & \textbf{-5}.15081  \\ \hline
%           &                 &           &           \\ 
           \hline
        & -0.749672       & \textbf{-0.749}256 & \textbf{-0.749}255 \\ %\hline
0.1           & -1.56062        & \textbf{-1.5}5665  & \textbf{-1.5}5665  \\ %\hline
           & -2.96323        & \textbf{-2.96}002  & \textbf{-2.96}002  \\ %\hline
           & -5.18772        & \textbf{-5}.09398  & \textbf{-5.1}1195  \\ \hline
%           &                 &           &           \\ 
           \hline
       & -0.758893       & \textbf{-0.75}7022 & \textbf{-0.75}7022 \\ %\hline
 0.05          & -1.56062        & \textbf{-1.56}447  & \textbf{-1.56}447  \\ %\hline
           & -2.97899        & \textbf{-2.978}42  & \textbf{-2.978}42  \\ %\hline
           & -5.21764        & \textbf{-5}.10019  & \textbf{-5}.10619  \\ \hline
%           &                 &           &           \\ 
           \hline
          & -0.763507       & \textbf{-0.76350}8 & \textbf{-0.76350}8 \\ %\hline
0           & -1.57201        & \textbf{-1.57201}  & \textbf{-1.5720}2  \\ %\hline
           & -2.98348        & \textbf{-2.983}52  & \textbf{-2.983}52  \\ %\hline
           & -5.10866        & \textbf{-5.1}3648  & \textbf{-5}.47278  \\ %\hline
\end{tabular}
\end{table}

%Use SI Units. Conventional (non-SI) quantities may follow parenthetically if the author desires.

\subsection{Sood Two Group Transport Problem}
The final problem we consider is a two group critical-slab problem  considers neutron transport in a two material, two region system consisting of a fuel region and a reflector region from \cite[Problem 59]{Sood:2003hl}. The original problem, being a $k$-eigenvalue benchmark, does not specify neutron speeds. We set the neutron speeds in the two groups to be $1$ and $10$ cm/$\mu$s. The system is designed to be critical, and the right-most eigenvalue is expected to be non-zero, but small, as a consequence of the numerical solution. The ability of a data driven method to capture this extremely small, negative eigenvalue necessarily requires computing the solution out to a very long time. Instead of requiring a corresponding very large number of time steps, allowing a variable time step greatly reduces the computational and storage costs associated with these kinds of problems.

This problem was simulated using a $S_N$ approximations with varying angles, $100$ spatial cells, and $300$ time steps logarithmically spaced, having an initial time step of $\approx0.001$ seconds, and ending with a time step of $\approx0.1$ seconds. The largest real eigenvalues recovered by VDMD are tabulated against the  $\alpha$ eigenvalue estimated by the standard $k$-$\alpha$ iteration \cite{hill:1983wj}. We note that VDMD estimates other eigenvalues, while $k$-$\alpha$ iterations are limited to computing the rightmost eigenvalue in the complex plane. We are able to use $k$-$\alpha$ iterations for this problem because it is close to critical. If we reduced the slab size, the system would become far from critical, and make the $k$-$\alpha$ iteration procedure unworkable \cite{Ortega:2017tw,kornreich2005time}.

\begin{table}[]
\centering
\caption{\bf Comparison of Eigenvalues for the critical slab problem computed using $k$-$\alpha$ iterations and VDMD.}\label{tab:sood}
\begin{tabular}{l|ll}
Method & $k$-$\alpha$ Eigenvalue ($\mu \mathrm{s}^{-1}$) & VDMD Eigenvalue ($\mu \mathrm{s}^{-1}$)\\ \hline \hline & & \\ 
$S_8$        & -1.058231$\times 10^{-5}$       & -1.058209$\times 10^{-5 }$  \\ %\hline
$S_{16}$       & -4.265406$\times 10^{-6}$       & -4.265209$\times 10^{-6}$   \\ %\hline
$S_{64}$       & -2.373016$\times 10^{-6}$       & -2.373179$\times 10^{-6}$   \\ %\hline
$S_{128}$      & -2.281708$\times 10^{-6 }$      & -2.281636$\times 10^{-6}$  \\ %\hline
\end{tabular}
\end{table}

While the exact, dominant eigenvalue for this problem is 0 because we have numerical error in the numerical solution (due to spatial, time, and angular discretizations), the eigenvalues for a given numerical instantiation of the problem are expected to be ``close'' to zero. This is indeed what we observe in Table \ref{tab:sood}. All of the eigenvalues estimated are subcritical and near zero. Note that as the number of angles in the discrete ordinates calculation increase, the eigenvalues get closer to zero. The results from $k$-$\alpha$ iteration and from VDMD agree to 4 or more digits in all calculations. 

\section{CONCLUSIONS}

The variable dynamic mode decomposition (VDMD) is an extension to the DMD method that is  capable of determining time eigenvalues of the neutron transport operator as effectively as DMD but without the requirement of uniformly-space time steps. We believe that this method will be useful to analyze systems where time-dependent phenomena are important and a range of time scales is also present, as in common in neutronics problems due to the separation of scales of prompt and delayed neutrons. 

There are several directions for future exploration into VDMD. The VDMD approach is naturally suited to adaptive time stepping methods. One could use the approximate operator estimated by VDMD to predict the solution after the next time step and use this result in a time step controller. Additionally, because VDMD, and also DMD, can be applied to multiphysics problems (particularly those with thermal-hydraulics negative feedback), applications of it to estimate properties of the multiphysics system should be explored. Finally, how to apply VDMD to measurement data is an open question. Because measurement data is not generated by a time integration algorithm, VDMD is not directly applicable to these problems. It may be possible, however, to formulate the problem using time averages to use a similar approach.

%, and demonstrating its efficacy on such algorithms including non-backward Euler time integration should be explored in future work
 %Moreover, VDMD, as well as DMD, should be explored in multiphysics problems, where DMD and VDMD can be used to estimate evolving modes that do not exactly correspond to $\alpha$ eigenvalues. Work in this direct has been reported by \cite{di2020dynamic}. The VDMD approach should enable further analysis of these systems. 

\section*{Acknowledgments}

I.V.'s portion of this work was funded by the Center for Exascale Monte-Carlo Neutron Transport (CEMeNT) a Predictive Science Academic Alliance Program (PSAAP-III) project funded by the Department of Energy, grant number: DE-NA003967. E.S.\ was supported by Los Alamos National Laboratory under contract number: 633356/CW7080.

\pagebreak
\bibliographystyle{style/ans_js}                                                                           %custom ANS journal submission template bibliography style
\bibliography{bibliography}

\end{document}